\documentclass[english,twocolumn,superscriptaddress,
  longbibliography]{revtex4-1}

\usepackage{xcolor}
\usepackage{graphicx}
\usepackage{amsmath}
\usepackage{amsfonts}
\usepackage{color}
\usepackage{babel}
\usepackage{url}
\usepackage{hyperref}
\usepackage{tikz}
\usepackage{CJKutf8}

\definecolor{lime}{HTML}{A6CE39}
\DeclareRobustCommand{\orcidicon}{
	\begin{tikzpicture}
	\draw[lime, fill=lime] (0,0) 
	circle [radius=0.16] 
	node[white] {{\fontfamily{qag}\selectfont \tiny ID}};
	\draw[white, fill=white] (-0.0625,0.095) 
	circle [radius=0.007];
	\end{tikzpicture}
	\hspace{-2mm}
}

\foreach \x in {A, ..., Z}{\expandafter\xdef\csname
  orcid\x\endcsname{\noexpand\href{https://orcid.org/\csname
      orcidauthor\x\endcsname} {\noexpand\orcidicon}} }

\begin{document}

\title{Designing linear lattices for round beam in electron storage rings
  using the solution by linear matrices analysis}
 
\author{Yongjun Li (\begin{CJK}{UTF8}{gbsn}李永军\end{CJK})\orcidA{}}\thanks{email: yli@bnl.gov}
\affiliation{Brookhaven
  National Laboratory, Upton 11973, New York, USA}

\author{Robert Rainer\orcidB{}} \affiliation{Brookhaven National Laboratory,
  Upton 11973, New York, USA}
 
\begin{abstract}
  For some synchrotron light source beamline applications, a round beam is
  preferable to a flat one. A conventional method of obtaining round beam
  in an electron storage ring is to shift its tune close to a linear
  difference resonance. The linearly coupled beam dynamics is analyzed
  with perturbation theories, which have certain limitations. In this
  paper, we adopt the Solution by LInear Matrices (SLIM) analysis to
  calculate exact beam sizes to design round beam lattices. The SLIM
  analysis can deal with a generally linearly coupled accelerator
  lattice. The effects of various coupling sources on beam emittances and
  sizes can be studied within a self-consistent frame. Both the on- and
  off-resonance schemes to obtain round beams are explained with
  examples. The SLIM formalism for two widely used magnet models:
  combined-function bending magnets, and planar wigglers and undulators,
  is also derived.
\end{abstract}
 
\maketitle

\section{\label{sect:intro}introduction}

  Round beam rather than a flat one is preferable for some beamline
  applications in the synchrotron light source community. Concurrently, an
  increased vertical beam size can significantly improve beam lifetime as
  well, particularly in extremely low emittance rings. Therefore, some
  future diffraction-limited light sources, such as
  ALS-U~\cite{Steier2018} and APS-U~\cite{Borland2017}, are planning to
  operate with a round beam mode.  Most of light source rings only have
  horizontal bending magnets, which leads to an intrinsically flat
  beam. Either dedicated devices, such as skew quadrupoles, or some
  imperfections in magnets, such as normal quadrupole roll errors, can
  couple the beam motion transversely. Conventionally, a geometric round
  beam in an electron machine is obtained by: (1) equally distributing the
  natural horizontal emittance into the horizontal and vertical planes
  $\epsilon_x=\epsilon_y$ through shifting the machine's tune close to a
  linear difference resonance $\nu_x-\nu_y-n=0$, with $n$ an integer, (2)
  adjusting the envelop Twiss functions so that $\beta_x=\beta_y$ at the
  locations of radiators. Here we also assume that radiators are located
  at non-dispersive sections, because achromat lattices are often adopted
  for light source rings. The beam emittances and sizes for this
  on-resonance coupling case were often analyzed with perturbation
  theories, such as~\cite{Bryant1975,Guignard1979,Franchi2011}
  etc. However, when the linear coupling is sufficiently strong, such
  perturbation analyses might not be accurate any longer and a more
  accurate analysis might be considered necessary.

  In the presence of linear coupling, the uncoupled 2-dimensional
  Courant-Synder parameterization~\cite{Courant1958} can be generalized to
  the 4-dimensional coupled motion. Such parameterizations, proposed by
  Ripken and his colleagues~\cite{Borchardt1988, Willeke1989} and further
  developed by Lebedev and Bogacz~\cite{Lebedev2010} are already
  available. There are also some other exact
  parameterizations~\cite{Edwards1973,luo2004,wolski2006}. These analyses
  only deal with linear Hamiltonian systems, the radiation damping and
  quantum excitation diffusion for electron beams are not
  considered. Therefore, the equilibrium emittance for electron storage
  rings has not been derived here. Instead, the following emittance
  re-distribution approximation~\cite{Guignard1979},
  \begin{equation}\label{eq:emitPert}
    \epsilon_x=\frac{1+2k^2}{1+4k^2}\epsilon_{x,0},\;
    \epsilon_y=\frac{2k^2}{1+4k^2}\epsilon_{x,0}
  \end{equation}
  is often used. Here $k=\frac{|\kappa|}{\Delta\nu}$, $\kappa$ is the
  well-known coupling coefficient given in
  ref.~\cite{Guignard1979,Guignard1976}, $\Delta\nu=\nu_x-\nu_y-p$ is the
  distance from the resonance, $\epsilon_{x,0}$ is the horizontal
  emittance for the uncoupled motion, and the natural vertical emittance
  $\epsilon_{y,0}$ is negligible.  Eq.~\eqref{eq:emitPert} is only valid
  by assuming: (i) coupling coefficient $\kappa$ are sufficiently weak to
  be considered as perturbations, (ii) the total transverse emittance
  remains as a constant, and (iii) the coupling is caused by a single
  isolated resonance. (iv) the vertical dispersion is negligible. Exact
  computations as shown later in this paper indicate that the
  approximation in Eq.~\eqref{eq:emitPert} can break down when these
  assumptions are violated, particularly when vertical dispersion is blown
  up.
  
  In this paper, to design round beam lattices for light source rings, we
  adopt an exact and self-consistent analysis -- the Solution by LInear
  Matrices (SLIM) technique, developed by Chao back in the
  1970--1980s~\cite{Chao1979, Chao1981, Chao2009}. This analysis can yield
  fruitful results such as the trajectory of the electron distribution
  center and the beam sizes and shapes in phase space. Linear coupling
  effects among the horizontal, vertical, and longitudinal motions are
  included in a straightforward manner even without introducing the
  auxiliary Twiss functions. An alternate, and also exact approach is to
  calculate the coupled synchrotron-radiation integrals~\cite{Ohmi1994},
  which was implemented in the code \textsc{SAD}~\cite{hirata1988} and
  \textsc{AT}~\cite{terebilo2001}, which could also be used for this
  purpose. We used \textsc{AT} and SLIM to compute a same coupled NSLS-II
  lattice and confirmed that their emittance computations are equivalent.

  The remainder of this paper is outlined as follows:
  Sect.~\ref{sect:slim} briefly explains the principle of the SLIM
  analysis. Sect.~\ref{sect:errors} introduces the linear couplings due to
  two random errors: Subsect.~\ref{sect:orbit} reviews the methods to take
  closed orbit errors into account; and the effect of normal quadrupole
  roll errors is investigated in Subsect.~\ref{sect:quadroll}.
  Sect.~\ref{sect:design} explains two round beam schemes, i.e., on- and
  off-resonance optics using examples. In Sect.~\ref{sect:operation}, we
  explain the strategies on the orbit and optics correction for different
  coupled lattice schemes, and indicate the direct tracking-based dynamic
  aperture optimization can still be used for nonlinear dynamics
  design. Some discussion and a brief summary is given in
  Sect.~\ref{sect:summary}. The formalism of two commonly used magnetic
  devices, combined-function bending magnets and planar wigglers and
  undulators, are given in the Appendix~\ref{sect:A-I} and \ref{sect:A-II}
  respectively.

\section{\label{sect:slim}SLIM and auxiliary Twiss functions}

  The detailed SLIM formalism can be found in the
  references~\cite{Chao1979, Chao1981, Chao2009, ChaoNote} and has been
  implemented in some accelerator codes, such as MAD-X~\cite{madx}. Here
  we will briefly explain its basic principles. It deals with the motion
  of a charged particle in a linear electromagnetic device by purely using
  their transport matrices. First, symplectic one-turn linear matrices for
  a storage ring are used to compute the eigenvalues and eigenvectors. The
  eigenvalues indicate whether the linear motion is stable or not, and
  provides the fractional parts of the tunes when the motion is
  stable. The eigenvectors evolving along the ring provide information
  about closed orbit, beam size, etc. For electron rings, non-symplectic
  one-turn matrices including the radiation damping are used to compute
  the damping rates. Then equilibrium emittances are obtained by balancing
  the quantum diffusion and radiation damping in all radiating magnets
  around the ring. The particle distributions within a bunch along the
  ring can be given with 21 independent second moments. In the presence of
  linear couplings, no approximation is needed and therefore, the
  computations are exact. The ring's global emittances and the local
  $s$-dependent beam sizes are derived within a self-consistent frame.

\begin{figure}[!ht]
    \centering
    \includegraphics[width=\columnwidth]{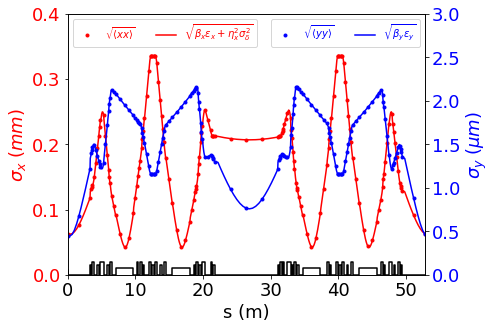}
    \caption{\label{fig:nsls2dba} NSLS-II DBA lattice magnet layout in a
      supercell. The beam sizes are represented with the 2nd
      moments in the SLIM analysis (dots) and the Courant-Snyder
      parameters (lines) for the uncoupled motion agrees with each
      other. Note that the vertical axes' scales and units for the
      horizontal (left) and vertical (right) beam sizes are different.}
  \end{figure}

  When there is no linear coupling, two SLIM's second moments
  $\left<xx\right>,\left<yy\right>$ were confirmed to agree with the beam
  sizes obtained with Sands's formalism~\cite{Sands1970} using the
  Courant-Snyder parameterization.  Throughout this paper the National
  Synchrotron Light Source II (NSLS-II) double bend achromat (DBA) bare
  lattice was used for the proof of concept. Fig.~\ref{fig:nsls2dba}
  shows the magnet layout of its one supercell composed of two mirror
  symmetrical DBA cells, and the uncoupled beam sizes along $s$, computed
  using these two  methods.

\begin{figure}[!ht]
    \centering
    \includegraphics[width=\columnwidth]{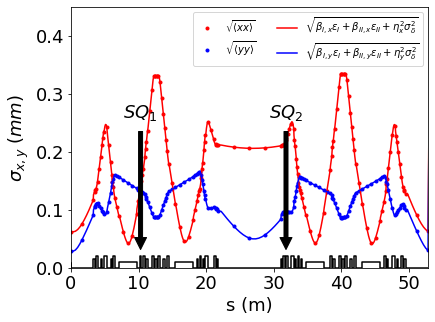}
    \caption{\label{fig:slimRM} Beam sizes obtained with the SLIM analysis
      (dots) and Ripken parameterization (line) in the presence of linear
      coupling. The locations of two skew quadrupoles are marked. By
      assigning them the same $|K_1|$ value, but alternating polarities
      along the ring, the vertical emittance and overall beam sizes will
      blow up. The dotted lines are the second moments obtained with the
      SLIM analysis, and the solid lines are computed with the coupled
      Twiss functions in Ripken parameterization. Nevertheless, for both
      cases, the equilibrium emittances $\epsilon_{I,II}$ and energy
      spread $\sigma_{\delta}$ need to be computed with the SLIM analysis
      first.}
  \end{figure}

\begin{figure}[!ht]
    \centering \includegraphics[width=\columnwidth]{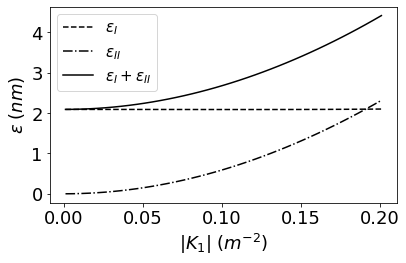}
    \caption{\label{fig:emitVsSQ} The correlation between the transverse
      emittances and the skew quadrupoles' strength $|K_1|$. The vertical
      emittance (the dot-dash line) grows faster than the reduction of the
      horizontal emittance (the dashed line). The total emittance is not
      constant when the coupling becomes strong, which leads to the
      breakdown of Eq.~\eqref{eq:emitPert}.}
  \end{figure}

  Linear coupling can then be introduced into the lattice by inserting two
  $0.2\;m$ long skew quadrupoles inside each supercell as shown in
  Fig.~\ref{fig:slimRM}. The first family of skew quadrupoles ($SQ_1$) are
  located within the achromat dispersive sections, while the second family
  ($SQ_2$) are located at the original dispersion-free sections. As a
  numerical experiment, all 30 skew quads were set with the same
  normalized focusing value $|K_1|=\left|\frac{1}{B\rho}\frac{\partial
    B_y}{\partial y}\right|$ but given alternating polarities
  ($+,\;-,\;+\;\cdots$) along the ring. Here $B\rho$ is the beam
  rigidity. The variation of the emittances with $|K_1|$ can then be
  computed with the SLIM analysis as shown in Fig.~\ref{fig:emitVsSQ}. The
  vertical emittance is observed to grow fast with the coupling strength,
  which leads to the sum of two emittances becoming variable and no longer
  constant. Eq.~\eqref{eq:emitPert} is only valid when the coupling
  coefficient $\kappa$ is sufficiently small.

  No auxiliary Twiss functions are needed in the SLIM analysis as the
  particle distributions are obtained by computing 21 independent second
  moments projected on the corresponding subspaces. However, it is worth
  noting are these coupled Twiss functions parameterized with Ripken
  method~\cite{Borchardt1988, Willeke1989} are still useful in
  interpreting the same physics meanings.  Given an equilibrium emittances
  $\epsilon_{x,y}$ and energy spread $\sigma_{\delta}$, the beam size
  along the ring can also be computed with the following
  formula~\cite{sagan1999} as illustrated in Fig.~\ref{fig:slimRM}.
  \begin{equation}\label{eq:sigma4b}
    \sigma_{x,y}^2 = \beta_{I,(x,y)}\epsilon_I+\beta_{II,(x,y)}\epsilon_{II}+
    \eta_{x,y}^2\sigma_\delta^2.
  \end{equation}

\begin{figure}[!ht]
    \centering \includegraphics[width=\columnwidth]{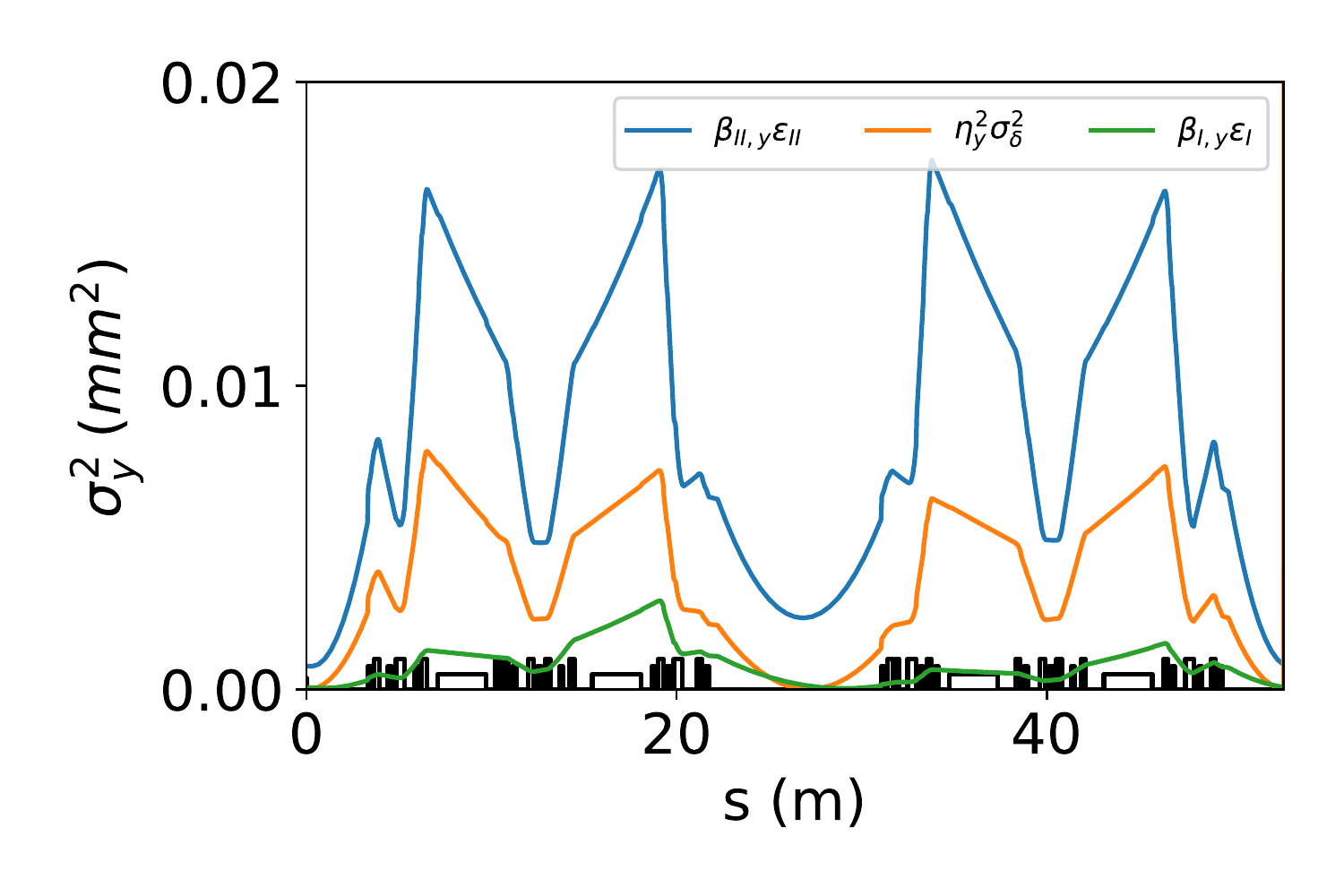}
    \caption{\label{fig:sigmaDecomp} Vertical beam size decomposition:
      coupled vertical dispersion (1) generates a considerable amount of
      mode II emittance contribution, which contributes
      $\beta_{II,y}\epsilon_{II}$ (blue), (2) introduces local energy
      oscillation, which contributes $\eta^2_y\sigma^2_{\delta}$ (yellow);
      (3) the coupled $\beta_{I,y}$ function couples some mode I
      emittance, which contributes $\beta_{I,y}\epsilon_{I}$ (green).}
  \end{figure}

  Using Eq.~\eqref{eq:sigma4b}, we can further understand the blow-up of
  vertical beam size by decomposing it into three components as shown in
  Fig.~\ref{fig:sigmaDecomp}.  When the dispersion is coupled from the
  horizontal plane to the vertical plane, it generates a considerable
  amount of mode II emittance $\epsilon_{II}$ and introduces local
  vertical energy oscillation $\eta_y\sigma_{\delta}$ as well. In the
  meantime, the coupled $\beta_{I,y}$ function can also generate a
  contribution $\beta_{I,y}\epsilon_{I}$.  When the vertical dispersion is
  sufficiently large, even no significant beam size change is observed in
  the horizontal plane, a new equilibrium state is formed in the vertical
  plane.

\section{\label{sect:errors} Coupling from random errors}  
  
  Linear coupling can be from not only dedicated magnets, such as skew
  quadrupoles and solenoids, but some random errors. In this section, the
  effects of two primary sources, the vertical closed orbit through
  sextupoles and the roll errors of normal quadrupoles, are discussed.

  \subsection{\label{sect:orbit}Closed orbit errors}

  In reality, small misalignments and magnet imperfections are
  unavoidable. Therefore, when circulating beam reaches an equilibrium
  distribution around a closed orbit, it always differs, at least
  slightly, from the design orbit. In this section, the technique for
  studying the effect of closed orbit on beam size is briefly
  explained. Closed orbit, if it exists, can be accurately obtained by
  performing an iterative one-turn 6-dimensional tracking till a
  convergence is reached. This method is widely used in many existing
  lattice codes, such as \textsc{elegant}~\cite{Borland2000} and
  \textsc{mad}~\cite{Grote1989}. Alternatively, the SLIM analysis adds a
  seventh component, which is always a unitary constant $1$, to expand
  one-turn transport matrices to a $7\times7$ format. The upper-left
  $6\times6$ corners of these matrices are the usual transport
  matrices. The $7^{th}$ row of all M matrices is always filled with
  $[0\;0\;0\;0\;0\;0\;1]$. The remaining upper-right $6\times1$ is the cause
  of orbit distortion, such as dipole errors.  In this case, the closed
  orbit corresponds to the eigenvectors with eigenvalues of $1$. The
  nonlinear kicks from nonlinear multipoles can also be accounted for by
  iteratively updating the transport matrices around the local closed
  orbit to reach a convergence.

  Once the closed orbit is obtained, the transport matrix of each magnet
  needs to be updated with the reference to it. Thick-lens off-axis
  transport matrices can be computed in different ways. A simple
  kick-drift model is often used to obtain an approximation of thick
  elements. First, a thick element is sliced into multiple pieces, e.g.,
  $n$ pieces. Each piece is approximated by two components of drift with a
  thin-lens kick in-between. By concatenating those slices sequentially, a
  thick model is obtained. For example, a thick lens sextupole can be
  approximated as the sequential products of the following kick-drift
  matrices,
  \begin{equation}\label{eq:sextCo} 
    R_{co,i} = M_dM_kM_d,
  \end{equation}
  with
  \begin{equation}
    M_d =
    \begin{bmatrix}
     1 & \frac{l_s}{2} & 0 & 0 & 0 & 0 \\
     0 & 1 & 0 & 0 & 0 & 0 \\
     0 & 0 & 1 & \frac{l_s}{2} & 0 & 0 \\
     0 & 0 & 0 & 1 & 0 & 0 \\
     0 & 0 & 0 & 0 & 1 & \frac{l_s}{2\gamma^2} \\
     0 & 0 & 0 & 0 & 0 & 1
    \end{bmatrix}, \nonumber
  \end{equation}
  and
  \begin{equation}
    M_k = 
    \begin{bmatrix}
     1 & 0 & 0 & 0 & 0 & 0 \\
     -\lambda x_{0,i} & 1 & \lambda y_{0,i} & 0 & 0 & \frac{\lambda }{2}(x_{0,i}^2-y_{0,i}^2) \\
     0 & 0 & 1 & 0 & 0 & 0 \\
     \lambda y_{0,i} & 0 & \lambda x_{0,i} & 1 & 0 & -\lambda x_{0,i}y_{0,i} \\
     -\frac{\lambda }{2}(x_{0,i}^2-y_{0,i}^2) & 0 & \lambda x_{0,i}y_{0,i} & 0 & 1 & 0 \\
     0 & 0 & 0 & 0 & 0 & 1
    \end{bmatrix}. \nonumber
  \end{equation}
  Here $R_{co,i}$ stands for the $i^{th}$ slice's matrix around the local
  closed orbit coordinates $(x_{0,i},y_{0,i})$,
  $\lambda=\frac{l_s}{B\rho}\frac{\partial^2B_y}{\partial^2x}$ is the
  normalized focusing integral of this slice, and $\gamma$ is the
  relativistic factor. When each slice's $l_s$ is reasonably small, the
  thick-lens $R_{co}=R_{co,n}\cdots R_{co,2}R_{co,1}$ will be sufficiently
  accurate. Since the kick-drift scheme is adopted, such thick-lens
  matrices are automatically symplectic. This $2^{nd}$-order thick-lens
  model has been adopted in our investigation. More complicated high order
  kick-drift schemes, as explained in ref.~\cite{Yoshida1990}, are also
  available.

  Another method to obtain a thick element's off-axis transport matrix is
  to compute its higher order on-axis matrices first, then the following
  truncated Taylor map~\cite{Carey1998} is used,
  \begin{equation}
    R_{co} = R + 2TX_0 + 3UX_0^2 + \cdots,
  \end{equation}
  where $X_0$ represents the 6-dimensional closed orbit vector at the
  magnet entrance and $R,T,U$ are the $1^{st}-3^{rd}$ order on-axis
  transport matrices which can be obtained with the truncated power series
  algorithm~\cite{Berz1988}. However, the truncated map is usually not
  perfectly symplectic, and can introduce an artificial damping or
  excitation. A symplectification process might be considered, if needed.

\begin{figure}[!ht]
    \centering
    \includegraphics[width=\columnwidth]{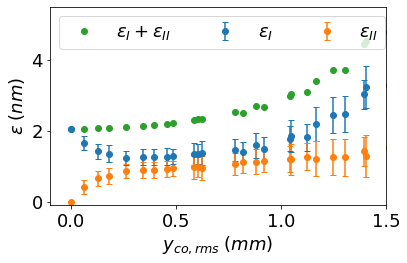}
    \caption{\label{fig:emVsCo} Correlation between the emittances and RMS
      vertical orbit distortions $y_{co,rms}$. With gradually increasing
      orbit distortion, a very flat beam becomes a near-round one. When
      the closed orbit distortion is sufficiently large, it becomes flat
      (and tilted) but with larger emittances. The total emittance is
      observed to be blown up, albeit rather slowly.}
  \end{figure}

  From Eq.~\eqref{eq:sextCo}, a vertical offset through sextupoles
  introduces linear coupling (non-zero $m_{23}$ and $m_{41}$). In modern
  high brightness light source rings, strong sextupoles are needed to
  correct chromaticity and enlarge dynamic aperture. Small closed orbit
  errors might introduce some coupling which can blow up the vertical
  emittance as shown in Fig.~\ref{fig:emVsCo}. The summation of transverse
  emittances increases gradually with the amplitude of vertical orbit. The
  idea of the constant emittance assumption from Eq.~\eqref{eq:emitPert}
  appears valid only when the root-mean-square (RMS) closed orbit error is
  less than about $200\mu m$ for the NSLS-II storage ring.

  The above computation is carried out when the machine's fractional tune
  (0.22/0.26) is not near the difference resonance. When the uncoupled
  tune is close to the difference resonance $\nu_x-\nu_y=n$, the vertical
  beam emittance can be easily increased as will be discussed in
  Sect.~\ref{subsect:onres}.
  
  \subsection{\label{sect:quadroll}Normal quadrupole roll errors}

\begin{figure}[!ht]
    \centering
    \includegraphics[width=\columnwidth]{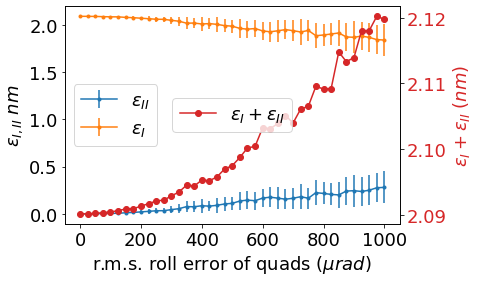}
    \caption{\label{fig:emitVsQR_offRes} Emittance variation with
      quadrupole roll errors when the machine's tune (0.22/0.26) is off a
      difference resonance.}
  \end{figure}

  Another linear coupling source is from the random normal quadrupole roll
  errors. Quadrupoles can be aligned within several hundreds of
  microradians roll angles using the modern alignment techniques. If the
  linear tune is sufficiently separated from resonances, even though the
  total beam emittances are only slightly increased (about 1-2\%), a
  significant part of the transverse emittance can be gradually
  redistributed to the vertical plane, as shown in
  Fig.~\ref{fig:emitVsQR_offRes}.  While the machine's tune is
  sufficiently close to a difference resonance, even small roll errors can
  easily couple the transverse motion. This is the most common way to
  obtain an approximately round beam, Although it can be explained well
  with perturbation theory, this particular case will be re-analyzed
  exactly with the SLIM technique in Sect.~\ref{subsect:onres}.

\section{\label{sect:design}Two round beam schemes}

  Thus far, we have explained how to compute exact beam emittances and
  sizes in the presence of various coupling sources using the SLIM
  analysis. To obtain round beam in an electron storage ring, its vertical
  emittance needs to be blown up with either dedicated coupling elements
  (skew quadrupoles, solenoids), or by shifting the machine's tunes close
  to a difference resonance, and letting random coupling errors (such as
  quadrupole roll errors) to couple the emittances transversely. Below we
  quantitatively investigate two schemes with examples.

  \subsection{\label{subsect:onres}Round beam with on-resonance tune}

    Currently, the most common method to obtain round beam is by shifting
    the machine's tune close to a difference resonance.  With this
    on-resonance scheme, random imperfections are usually sufficient to
    couple the transverse motion. Because orbit displacements through
    sextupoles centers can be well controlled using the
    beam-based-alignment technique~\cite{lavine1988, portmann1995},
    quadrupoles' roll errors are regarded as the primary coupling sources,
    which are often at a level of several hundred microradians.

    Although this on-resonance scheme can be explained with the
    perturbation theory, we re-investigated it with the exact SLIM
    analysis. Our design goal is to make the beam to be round at the short
    straight centers (SSC). First, the local quadrupoles $QL_{1-3}$ there
    were re-tuned to let the local eigenvectors absolute values to be
    close (or the coupled $\beta_{(x,y),(I,II)}$ functions to be close if
    one prefers to use Eq.~\eqref{eq:sigma4b} instead.) Then the
    quadrupoles in the long straight sections $QH_{1-3}$ were tuned to
    shift the fractional tune close to a difference resonance. Here we
    assumed the RMS roll angles for quadrupoles is 500 $\mu rad$, which
    can be easily achieved with the current alignment technique. The
    needed adjustments on quadrupoles settings as listed in
    Tab.~\ref{tab:nsls2rb_onRes} can switch the nominal lattice to an
    on-resonance round beam lattice. The corresponding beam emittances,
    fractional tune and beam sizes (at SCCs) are also listed there.

    \begin{table}[bht]
      \caption{Parameters from flat beam to round beam with on-resonance tune}
      \centering
      \begin{tabular}{l p{2.3cm} p{2.3cm} r}
        \hline
        name & original value & new value & unit \\ [0.5ex] 
        \hline
        $K_{1,QL_1}$ & -1.61785 & -1.39864 & $m^{-2}$ \\ 
        $K_{1,QL_2}$ & 1.76477 & 1.73601  & $m^{-2}$ \\
        $K_{1,QL_3}$ & -1.51868 & -1.58219 & $m^{-2}$ \\
        $K_{1,QH_1}$ & -0.64196 & -0.62274 & $m^{-2}$ \\
        $K_{1,QH_2}$ & 1.43673 & 1.42475 & $m^{-2}$ \\
        $K_{1,QH_3}$ & -1.75355 & -1.76958 & $m^{-2}$ \\
        R.M.S. quad. roll  & 0 & 500 & $\mu rad$ \\
        $\epsilon_I/\epsilon_{II}$ & 2.096/0.0002 & 1.185/0.925& $nm$ \\ 
        $\nu_I/\nu_{II}$ \footnote{fractional part}& 0.22/0.26 &
        \bf{0.23}/\bf{0.23} & 1\\ 
        $\sigma_x/\sigma_y$ \footnote{averaged beam sizes observed
        at short straight centers} & 62.08/0.45 & \bf{43.1}/\bf{44.2} &
          $\mu m$ \\ \hline
      \end{tabular}
      \label{tab:nsls2rb_onRes}
    \end{table}

\begin{figure}[!ht]
      \centering
      \includegraphics[width=\columnwidth]{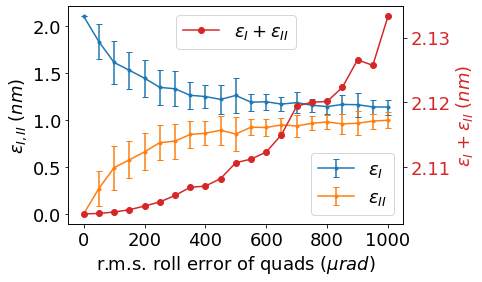}
      \caption{\label{fig:emitOnRes} On-resonance scheme: correlation
        between the beam eigen-emittances with r.m.s. quadrupole roll
        errors when the ideal machine's tune is on a difference
        resonance. Each error bar is the standard deviation for 50 random
        seeds.}
    \end{figure}

\begin{figure}[!ht]
      \centering
      \includegraphics[width=\columnwidth]{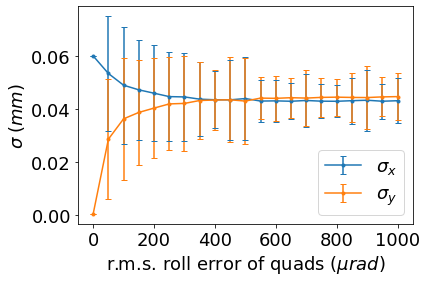}
      \caption{\label{fig:sigmaOnRes} On-resonance scheme: correlation
        between the quadrupole roll errors and beam sizes at short
        straight centers when the ideal machine's tune is on a difference
        resonance. The error bar is the standard deviation of 50 random
        seeds.}
    \end{figure}

\begin{figure}[!ht]
      \centering
      \includegraphics[width=\columnwidth]{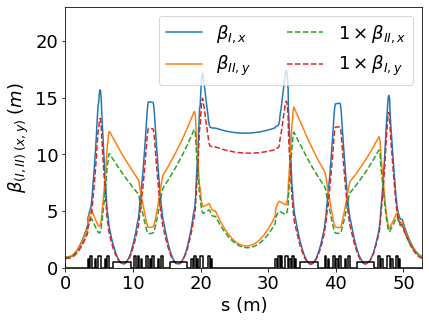}
      \caption{\label{fig:coupledBetaOnRes} On-resonance scheme: four
        coupled $\beta$ functions of one supercell for one random seed
        with $500\mu rad$ quadrupole roll errors. The local
        $\beta_{(I,II),x}$ at short straight sections are tuned close to
        $\beta_{(I,II),y}$, to obtain a geometric round beam there.}
    \end{figure}

\begin{figure}[!ht]
      \centering
      \includegraphics[width=\columnwidth]{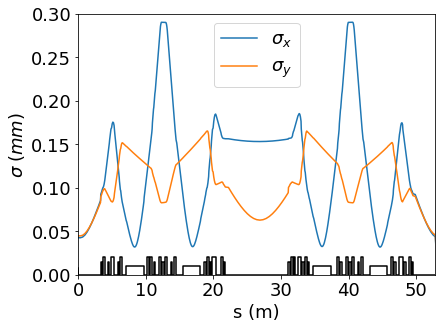}
      \caption{\label{fig:coupledSizeOnRes} On-resonance scheme:
        horizontal and vertical beam sizes of one supercell for one
        specific random seed with $500\mu rad$ quadrupole roll
        errors. Such data can be used for the beam lifetime estimation.}
    \end{figure}

    The eigen-emittances and beam sizes' variation with the quadrupole
    roll angles are illustrated in Fig.~\ref{fig:emitOnRes} and
    Fig.~\ref{fig:sigmaOnRes} respectively. Because quadrupole roll errors
    are random, the corresponding beam emittances and sizes have some
    statistical fluctuations as represented by error bars there. These
    figures illustrate the needed quadrupoles roll errors to get a
    geometric round beam.  With the quadrupole settings in
    Tab.~\ref{tab:nsls2rb_onRes} , a RMS error greater than $400\;\mu rad$
    will be sufficient to result in $\sigma_x\approx\sigma_y$. It is
    interesting to note that, the round beam size ($\sim44\;\mu m$) is
    significantly greater than the half of the uncoupled horizontal flat
    beam size ($\sim60\;\mu m$) as seen in Fig.~\ref{fig:sigmaOnRes}. It
    can be explained with Eq.~\eqref{eq:sigma4b}: although two mode's
    emittances $\epsilon_{I,II}$ are approximately equal, the sum of
    coupled $\beta_{I,(x,y)}+\beta_{II,(x,y)}$ is greater than the
    uncouple $\beta_{(x,y)}$ as illustrated in
    Fig.~\ref{fig:coupledBetaOnRes}. The beam sizes for one supercell with
    one specific random seed is illustrated in
    Fig.~\ref{fig:coupledSizeOnRes}, which can be used for the beam
    lifetime estimation.
       
    If quadrupoles are aligned accurately with smaller roll angles,
    eigen-emittances might not be equally distributed by the resonance
    coupling, i.e., $\epsilon_I>\epsilon_{II}$. In this case, we can still
    tune the local quadrupoles ($QL_{1-3}$) to make
    $\beta_{(I,II),x}<\beta_{(I,II),y}$ correspondly to get a geometric
    round beam.

\begin{figure}[!ht]
      \centering
      \includegraphics[width=\columnwidth]{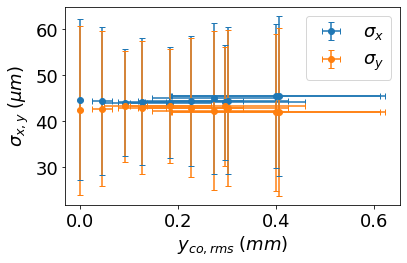}
      \caption{\label{fig:rollCo_onRes} On-resonance scheme: round beam
        profile can be maintained with gradually increased closed orbit
        errors when the machine's tune stays near a difference resonance.}
    \end{figure}

    Now we consider a more realistic situation by including the closed
    orbit distortion. Thus the skew quadrupole components due to the
    vertical offsets through sextupoles can be taken into account. A
    simulation shows that the round beam profiles can still be maintained
    when the closed orbit errors are within a quite wide range (see
    Fig.~\ref{fig:rollCo_onRes}). Once an even larger closed orbit
    distortion can drive the tune further away from the resonance,
    however, the beam profiles will gradually become flatter. Usually the
    closed orbit can be well controlled after the beam-based alignment
    technique is implemented, this error shouldn't be a concern.
    
  \subsection{Round beam with off-resonance tune}

    The second scheme to obtain round beam would be to use some strong
    skew/tilted quadrupoles (or solenoids), which would allow the
    machine’s tune to stay off resonances. In this case, the perturbation
    theory is not applicable. To obtain round beam at specific locations,
    the linear lattice design can be summarized as an optimization
    problem:
    \begin{enumerate}
      \item Assign some magnet's focusing strengths (such as normal and
        skew quadrupoles) $K_1$, and/or tilt angles $\phi$ as knobs.
      \item Simultaneously minimize:
        \begin{enumerate}
          \item $\epsilon_I$;
          \item $\epsilon_{II}$;
          \item $|\frac{\sigma_y}{\sigma_x}-1|$ at specific $s$ locations.
        \end{enumerate}
      \item subject to the following constraints:
        \begin{enumerate}
          \item the existence of stable linear solutions;
          \item the fractional tune is sufficiently away from low order
            resonances;
          \item $\epsilon_{I,II} < \epsilon_{threshold}$
          \item the tilt angle $|\theta_{xy}|<\theta_{threshold}$; 
          \item $K_1\in[K_{1,min},K_{1,max}]$;
          \item $|\frac{\sigma_y}{\sigma_x}-1|<d$ at specific $s$ locations;
        \end{enumerate}
    \end{enumerate}
    Here, the stable linear solution constraint means that three
    eigenvalues of the one-turn symplectic matrix must stay on the complex
    unity circle, $\epsilon_{threshold}$ is the emittance threshold to
    guarantee the required brightness,
    $\tan2\theta_{xy}=\frac{2\left<xy\right>}{\left<xx\right>-\left<yy\right>}$
    defines the tilt angle for the $x-y$ beam profile relative to the
    horizontal axis, $d$ represents an allowed tolerance for the beam to
    be a perfectly round shape.  This linear lattice design eventually
    needs to be optimized iteratively after taking the dynamic aperture and
    energy acceptance into account, but was not covered in this paper.

\begin{figure}[!ht]
      \centering
      \includegraphics[width=\columnwidth]{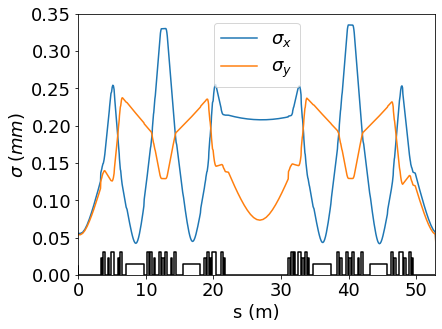}
      \caption{\label{fig:coupledSizeOffRes} Off-resonance scheme:
        horizontal and vertical beam sizes of one supercell.}
    \end{figure}

    A possible linear solution for the NSLS-II ring is listed in
    Tab.~\ref{tab:nsls2rb}. Three families of quadrupoles $QL_{1-3}$, and
    two families of skew quadrupoles $[SQ_1, SQ_2]$ are used. The beam
    size for one supercell is illustrated in
    Fig.~\ref{fig:coupledSizeOffRes}.

    \begin{table}[ht]
      \caption{Parameters from flat beam to round beam for off-resonance tune}
      \centering
      \begin{tabular}{l p{2.3cm} p{2.3cm} r} 
        \hline
        Name & original value & new value & unit \\ [0.5ex] 
        \hline
        $K_{1,QL_1}$ & -1.61785 & -1.23000 & $m^{-2}$ \\ 
        $K_{1,QL_2}$ & 1.76477 & 1.70791 & $m^{-2}$ \\
        $K_{1,QL_3}$ & -1.51868 & -1.61387 & $m^{-2}$ \\
        $K_{1,SQ1}$ & 0 & 0.085 & $m^{-2}$ \\ 
        $K_{1,SQ2}$ & 0 & -0.090 & $m^{-2}$ \\
        $\epsilon_I/\epsilon_{II}$ & 2.096/0.0002 & 2.095/1.460& $nm$ \\ 
        $\nu_I/\nu_{II}$ \footnote{fractional part} & 0.22/0.26 &
        \bf{0.59}/\bf{0.66} & 1\\ 
        $\sigma_x/\sigma_y $\footnote{observed at short straight centers}
        & 62.08/0.45 & \bf{55.5}/\bf{53.8} & $\mu m$ \\ 
        \hline
      \end{tabular}
      \label{tab:nsls2rb}
    \end{table}

\begin{figure}[!ht]
      \centering
      \includegraphics[width=\columnwidth]{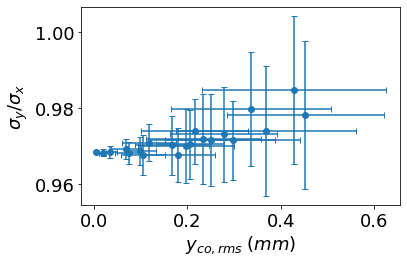}
      \caption{\label{fig:ratioVsCo} Off-resonance scheme: ratio of the
        vertical and horizontal beam sizes varies with the vertical closed
        orbit distortions. The initial ratio of transverse beam sizes is
        $\sigma_y/\sigma_x\approx0.97$, then a closed orbit that is
        increased gradually can slightly stretch the beam $x-y$ profile
        vertically. Each error bar represents the standard deviation of 20
        random seeds.}
    \end{figure}

    Once the imperfections of magnets and the misalignments are accounted
    for, beam will reach an equilibrium around a closed orbit with random
    quadrupole roll errors. A large quadrupole roll errors and closed
    orbit distortions could overstretch and tilt the round beams. For
    example, a simulation studying the deformation of a round beam due to
    vertical closed orbit distortions is shown in
    Fig.~\ref{fig:ratioVsCo}. For this particular lattice, since its tune
    is away from the difference resonance, the profile of the round beam
    is not noticeably sensitive to closed orbit distortions if they can be
    constrained to within $100-200\mu m$.

\begin{figure}[!ht]
      \centering
      \includegraphics[width=0.7\columnwidth]{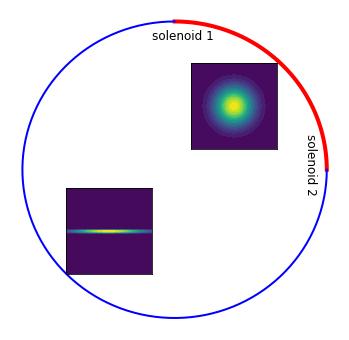}
      \caption{\label{fig:hybridMode} Schematic diagram of a hybrid
        flat-round beam mode in a storage ring. A pair of solenoids are
        used to form a local closed section with a round beam (red arc),
        while maintaining a flat beam in the rest part of the ring (blue
        arc.)}
    \end{figure}

    Recently a hybrid flat-round beam scheme~\cite{chao-deng2021} is being
    under investigation for a steady-state microbunching~\cite{ratner2010}
    ring. A similar idea has been studied using a different
    method~\cite{du2021} for a diffraction-limited light source ring. The
    schematic diagram of this hybrid flat-round beam mode for a storage
    ring is illustrated in Fig.~\ref{fig:hybridMode}, in which a pair of
    coupling elements are used to generate a local closed section with a
    round beam, while maintaining a flat beam in the rest part of the
    ring. Obviously this scheme is also an off-resonance scheme. Even a
    fully coupling beam is limited within a section, once there exist
    radiation elements, such as dipoles, inside, it changes the mode I
    emittance in the “flat beam” part of the ring, which can be exactly
    analyzed with the SLIM technique.

    It is worth to note that the off-resonance scheme is more robust to
    random errors than the on-resonance one by comparing
    Fig.~\ref{fig:rollCo_onRes} and \ref{fig:ratioVsCo}. The reason is
    simply because the on-resonance scheme solely relays on the random
    errors amplified by the linear resonance. While the off-resonance
    scheme takes advantage of strong and dedicated linear coupling magnets
    and the tune is far away from the resonance.

\section{\label{sect:operation}Orbit and optics correction, and dynamic aperture}

  In this section, we briefly discuss two topics related to linear
  lattice design, i.e., orbit and linear optics characterization and
  correction, and dynamic aperture optimization.
  
  \subsection{\label{correction}Orbit and linear optics control}

    Linear orbit and optics characterization and correction need to be
    implemented during beam commissioning and routine operation. For the
    on-resonance scheme, coupling sources are random, so that there is
    some uncertainty in real lattices, and no precise lattice models are
    available. The orbit response matrix for the various orbit control
    should be measured experimentally. The uncoupled linear optics could
    be characterized and corrected when the machine tune is sufficiently
    away the difference resonance, then some dedicated quadrupoles can be
    used to shift the tune close to the resonance.

    For the off-resonance scheme, coupling sources are well-defined
    magnetic elements integrated into the lattice model. Traditional
    methods, such as linear optics from closed orbit
    (LOCO)~\cite{safranek1997}, can be used to characterize the optics and
    identify needed corrections. The only difference is that the
    off-diagonal elements in orbit response matrix are not zeros any
    longer. Similarly coupled orbit response matrix should be used for
    various orbit controls as well.

  \subsection{\label{da}Dynamic aperture}
    
    Sufficient dynamic aperture (DA) and local momentum aperture (LMA) are
    always required for carrying out high efficiency injection and
    achieving desired beam lifetime. For the on-resonance scheme, DA and
    LMA should be optimized when the tune is sufficiently away the
    difference resonance. After some global quadrupole knobs are used to
    move the tunes to the linear difference resonance, tracking can be
    used to check the DA degradation. The linear chromaticities should be
    adjusted to an desired value for both modes with sextupoles. After DA
    tracking, statistical analysis of the DA are needed to provide the
    reduction percentile contours~\cite{Borland2017}. If a significant
    degradation of DA and/or LMA is found, some further optimization on
    uncoupled lattice is needed.

\begin{figure}[!ht]
      \centering
      \includegraphics[width=0.7\columnwidth]{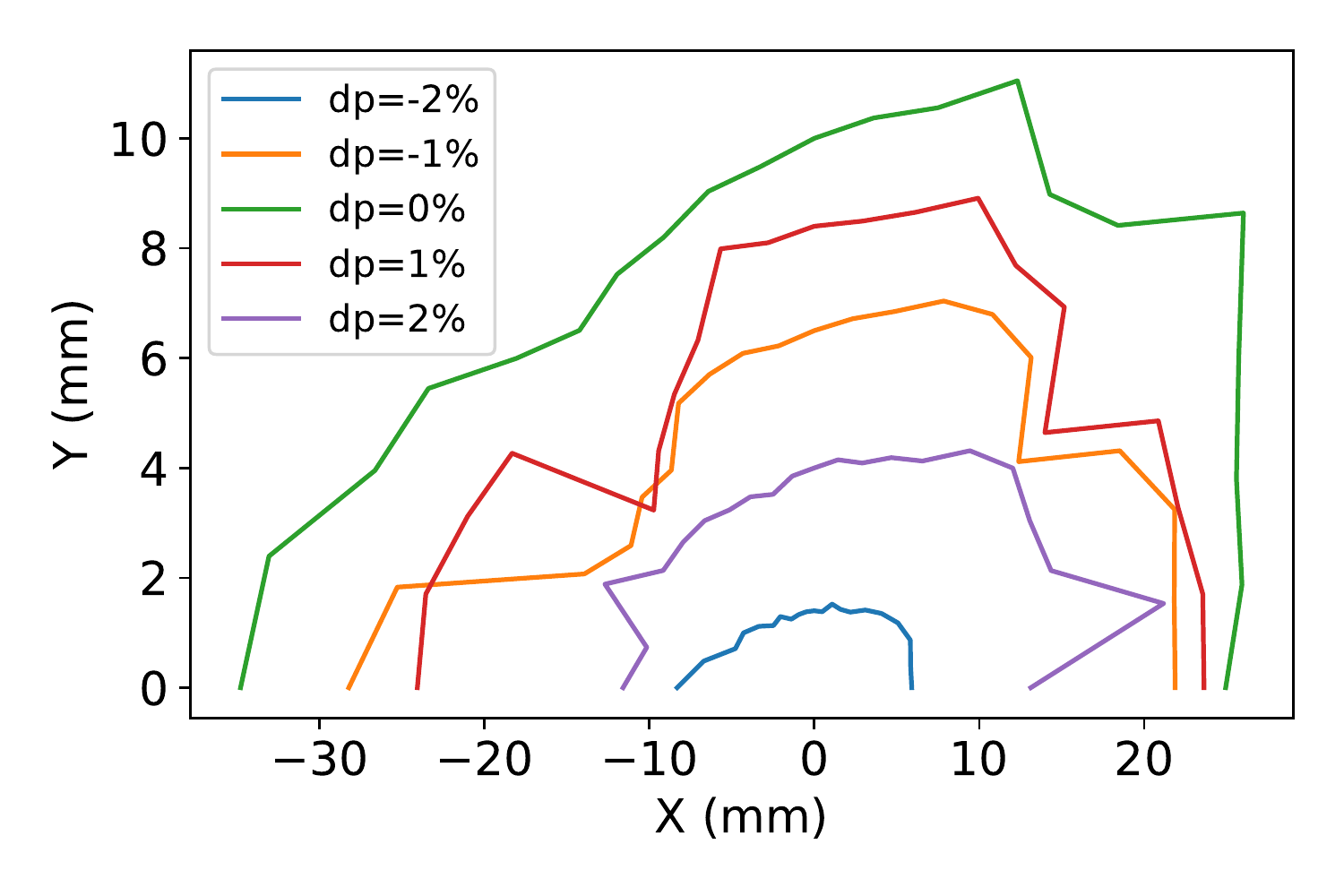}
      \caption{\label{fig:offResDA} Optimized on- and off-momentum dynamic
        apertures for the off-resonance scheme as listed in
        Tab.~\ref{tab:nsls2rb}.}
    \end{figure}

    For the off-resonance scheme, given a linear optics, direct
    tracking-based DA optimization techniques, such
    as~\cite{borland2009,yang2011,huang2014}, can be used to optimize
    nonlinear magnets configuration. First, the chromaticities should be
    adjusted to an desired value for both modes with sextupoles. Due to
    the existence of vertical dispersion, most sextupoles can contribute
    to the linear chromaticities. In our off-resonance scheme example, we
    used the least square method to correct the linear
    chromaticities. Then the response matrix of chromaticity with respect
    to sextupoles was decomposed to get its null vectors~\cite{huang2015}.
    Various linear combination of those null vectors were added on
    sextupoles to optimize DA and LMA without changing linear
    chromaticities. Sufficient DAs were obtained as illustrated in
    Fig.~\ref{fig:offResDA}. If the optimal solution can't satisfy the DA
    requirement, iteratively optimization for the linear optics should be
    considered necessary.

\section{\label{sect:summary}Summary}

  Two schemes to obtain round beam, i.e., with machine's tune sitting on-
  or off- difference resonance, are studied with the SLIM analysis. This
  exact analysis accounts for the linear coupling by calculating the
  quantum diffusion and radiation damping rate around a 6-dimensional
  closed orbit. The on-resonance scheme takes the advantage of the random
  imperfections to drive the beam to be coupled when the tune stays close
  to a difference resonance. This scheme is easy to implement in a real
  machine, however, beam profiles, coupled optics functions, and
  dispersions etc. have quite large and uncontrollable fluctuations. The
  off-resonance scheme can provide a more controllable and robust round
  beam, but needs to integrate dedicated magnets into the lattice to
  generate strong coupling.

  In modern light source storage rings, high-harmonic cavities are often
  used to stretch the bunch longitudinally. Under desired conditions, the
  local potential well basin is flattened. Strictly speaking, if a linear
  focusing is missing from any dimension, the SLIM analysis cannot be
  applied. In this case, the head and tail of the bunch are still focused
  by the nonlinear potential well, but the bunch center will coast because
  there is no linear focusing at $z=0$. To deal with this special case, an
  artificial, small linear RF focusing can be introduced. Although the
  bunches will have long longitudinal sizes, the transverse dynamics and
  energy spread computations are still correct because the longitudinal
  focusing is weak at most rings. However, this calculation will miss one
  class of effects, i.e., the nonlinear synchro-betatron coupling
  effects. When the longitudinal focusing comes purely from a nonlinear
  potential well, these nonlinear synchro-betatron effects could be
  significant. For cases like this, the SLIM analysis can still be used to
  design the linear lattice first, then detailed simulations are needed to
  study the nonlinear effects~\cite{chao-deng2021}.

  In real machine operations, maintaining round beam lattices can still be
  complicated and challenging. For example, when some users open or close
  their insertion device gaps, not only do the devices focusing properties
  change, but the equilibrium between the damping and quantum excitation
  changes simultaneously. Such changes are observed for all
  beamlines. Therefore, a feed-forward or feed-back system (or both) would
  be needed to stabilize machine's tune, coupling, local beam sizes
  etc. to account for these changes. As previously mentioned, the
  nonlinear beam dynamics is only mentioned lightly in this paper. Usually an
  iterative optimization between the linear and nonlinear lattice design
  would be needed once the SLIM analysis was deployed.
  
\begin{acknowledgments}
  We would like to thank Prof. Alexander W. Chao (Stanford Uni.) and
  Dr. Xiujie Deng (Tsinghua Uni.) for invaluable discussions, reading the
  origin manuscript and code-sharing, Dr. P. Kuske for pointing out a
  misinterpretation on the perturbation theory. The supports and
  discussion from the NSLS-II colleagues are greatly appreciated. This
  research used resources of the National Synchrotron Light Source II, a
  U.S. Department of Energy (DOE) Office of Science User Facility operated
  for the DOE Office of Science by Brookhaven National Laboratory under
  Contract No. DE-SC0012704.
\end{acknowledgments}

\appendix

\section{\label{sect:A-I}Combined-function bending magnets}
  Two magnet models commonly used in light source rings, combined-function
  bending magnets and wigglers and undulators, are not included in Chao's
  original literature. The appendices summarize their transport matrices,
  the corresponding changes when the radiation damping and closed orbit
  are considered. The quantum diffusion rate for planar wigglers and
  undulators are derived.

  The Hamiltonian of combined function bending magnet is given in
  ref.~\cite{Iselin1984}. Here a thick dipole is sliced into multiple
  pieces. The transport matrix around a closed orbit $x_0,y_0$ for one
  slice can be approximated as,
  \begin{equation}\label{eq:combBend}
    R_{co,i} = M_dM_kM_d,
  \end{equation}
  with
  \begin{equation}
    M_k =
    \begin{bmatrix}
     1 & 0 & 0 & 0 & 0 & 0 \\
     -K_hl_s & 1 & 0 & 0 & 0 & K_hl_sx_0+l_sh \\
     0 & 0 & 1     & 0 & 0 & 0 \\
     0 & 0 & K_1l_s & 1 & 0 & -K_1l_sx_0\\
     -K_hl_sx_0-l_sh & 0 & K_1l_sx_0 & 0 & 1 & 0 \\
     0 & 0 & 0 & 0 & 0 & 1
    \end{bmatrix}, \nonumber
  \end{equation}
  here $M_d$ is the drift's transport matrix as already given in
  Eq.~\eqref{eq:sextCo}, $K_h=K_1+h^2$ with $h=\frac{B_{y,0}}{B\rho}$ the
  reciprocal of bending radius. Correspondingly, after including the
  transverse gradient contribution, the radiation damping changes the
  diagonal $m_{66}$ element as,
  \begin{equation}
    m_{66} = 1-\frac{C_{\gamma}E^3}{\pi}
      \frac{l_s(B_x^2+B_y^2)}{(B\rho)^2}
  \end{equation}
  here $C_\gamma=8.846\times10^{-5} m/GeV^3$ for electrons, $B_y =
  B_{y,0}+\frac{\partial B_y}{\partial x}x_0$ and $B_x=\frac{\partial
    B_x}{\partial y}y_0$ are the vertical and horizontal field seen by
  electrons at a closed orbit $x_0, y_0$. When calculating closed orbit,
  nonvanishing elements in the $7^{th}$ column are
  \begin{align}
    &m_{67}=-\frac{C_{\gamma}E^3}{\pi}\frac{l_s(B_x^2+B_y^2)}{(B\rho)^2} \nonumber\\
    &m_{27}=\frac{m_{67}}{2}\frac{B_yl_s}{B\rho} \nonumber\\
    &m_{47}=\frac{m_{67}}{2}\frac{B_xl_s}{B\rho} \nonumber\\
    &m_{77}=1.
  \end{align}

\section{\label{sect:A-II}Planar insertion devices}

  Insertion devices such as wigglers and undulators are used as the main
  X-ray radiators on modern light source rings. In this appendix a planar
  Halbach type wiggler or undulator model~\cite{halbach1980} is integrated
  into the SLIM framework. The $s$-dependent vertical magnetic field of
  such a devices reads as,
  \begin{equation}
    B_y=\hat{B}_y\cosh(k_xx)\cosh(k_yy)\sin(ks),
  \end{equation}
  Here $k_x^2+k_y^2=k^2=(\frac{2\pi}{\lambda_l})^2$, $\lambda_l$ is the
  period length of the device, $\hat{B}_y$ is the peak field in the
  vertical plane. The linear focusing of such type devices can be
  extracted as explained in ref.~\cite{Smith1986}. If the magnetic poles
  are sufficiently wide, the weak focusing in the horizontal plane is
  negligible, i.e., $k_x\approx 0$, the nonvanishing linear transport
  matrix elements for one complete period are listed below.
  \begin{align}
    &m_{11} = m_{22} = m_{55} = m_{66} = 1 \nonumber\\
    &m_{12} = \lambda_l \nonumber\\
    &m_{33} = m_{44}  =  \cos(\sqrt{K_y}\lambda_l) \nonumber\\
    &m_{34} =  \frac{\sin(\sqrt{K_y}\lambda_l)}{\sqrt{K_y}} \nonumber\\
    &m_{43} = -\sqrt{K_y}\sin{\sqrt{K_y}\lambda_l} \nonumber\\
    &m_{56} = \lambda_l/\gamma^2.
  \end{align}
  Here $K_y=\frac{\hat{B}_y^2}{2(B\rho)^2}$. Note that a complete wiggler
  and undulator period is an achromat system, therefore,
  $m_{16}=m_{26}=0$. Other unlisted elements are all zeros. The radiation
  damping changes the diagonal element $m_{66}$ for one period as
  \begin{equation}
    m_{66} = 1-\frac{C_\gamma E^3\hat{B}_y^2\lambda}{2\pi(B\rho)^2}.
  \end{equation}
  The quantum diffusion rate for one period is
  \begin{equation}
    \left<|A_{\pm k}|^2\right> = 
    C_L\frac{\gamma^5|E_{k5}(s)|^2}{c\alpha_k}
    \frac{4\hat{B}_y^3\lambda_l}{3\pi(B\rho)^3},
  \end{equation}
  here $C_L=\frac{55}{48\sqrt{3}}\frac{r_e\hbar}{m_e}$, $\hbar$ is the
  reduced Planck’s constant, $r_e$ is the classical electron radius, $m_e$
  is the electron mass, $c$ is the speed of light, $\alpha_k$ is the
  radiation damping constant for mode $k=I,II,II$, $E_{k5}$ is the
  $5^{th}$ component of the normalized eigenvector for the mode $k$ at the
  location of this period. A relative small contributions from sideways
  photons can be included as well,
  \begin{equation}
    \left<|A_{\pm k}|^2\right>_{sideways}  = 
    C_L\frac{\gamma^3|E_{k(1,3)}(s)|^2}{c\alpha_k}
    \frac{2\hat{B}_y^3\lambda_l}{3\pi(B\rho)^3},
  \end{equation}
  Here $E_{k(1,3)}(s)$ is the $1^{st}$ or $3^{rd}$ component of the
  normalized $k^{th}$ mode's eigenvector in the horizontal or vertical
  plane. Here all photon emission events are assumed to be
  uncorrelated. The intrinsic dispersion effect~\cite{Zhang2021} is
  ignored because it is negligible comparing with bending magnets in
  the third and fourth generation light source rings.

  When calculating closed orbit, nonvanishing elements for one period in
  the $7^{th}$ column are
  \begin{align}
    &m_{67}=-\frac{C_\gamma E^3\hat{B}_y^2\lambda_l}{2\pi(B\rho)^2}\nonumber\\
    &m_{77}=1.
  \end{align}
  Unlike bending magnets, $m_{27}$ for a complete wiggler period is zero
  due to its alternating field polarities.
  
\bibliography{s4rb.bib}

\begin{thebibliography}{42}%
\makeatletter
\providecommand \@ifxundefined [1]{%
 \@ifx{#1\undefined}
}%
\providecommand \@ifnum [1]{%
 \ifnum #1\expandafter \@firstoftwo
 \else \expandafter \@secondoftwo
 \fi
}%
\providecommand \@ifx [1]{%
 \ifx #1\expandafter \@firstoftwo
 \else \expandafter \@secondoftwo
 \fi
}%
\providecommand \natexlab [1]{#1}%
\providecommand \enquote  [1]{``#1''}%
\providecommand \bibnamefont  [1]{#1}%
\providecommand \bibfnamefont [1]{#1}%
\providecommand \citenamefont [1]{#1}%
\providecommand \href@noop [0]{\@secondoftwo}%
\providecommand \href [0]{\begingroup \@sanitize@url \@href}%
\providecommand \@href[1]{\@@startlink{#1}\@@href}%
\providecommand \@@href[1]{\endgroup#1\@@endlink}%
\providecommand \@sanitize@url [0]{\catcode `\\12\catcode `\$12\catcode
  `\&12\catcode `\#12\catcode `\^12\catcode `\_12\catcode `\%12\relax}%
\providecommand \@@startlink[1]{}%
\providecommand \@@endlink[0]{}%
\providecommand \url  [0]{\begingroup\@sanitize@url \@url }%
\providecommand \@url [1]{\endgroup\@href {#1}{\urlprefix }}%
\providecommand \urlprefix  [0]{URL }%
\providecommand \Eprint [0]{\href }%
\providecommand \doibase [0]{http://dx.doi.org/}%
\providecommand \selectlanguage [0]{\@gobble}%
\providecommand \bibinfo  [0]{\@secondoftwo}%
\providecommand \bibfield  [0]{\@secondoftwo}%
\providecommand \translation [1]{[#1]}%
\providecommand \BibitemOpen [0]{}%
\providecommand \bibitemStop [0]{}%
\providecommand \bibitemNoStop [0]{.\EOS\space}%
\providecommand \EOS [0]{\spacefactor3000\relax}%
\providecommand \BibitemShut  [1]{\csname bibitem#1\endcsname}%
\let\auto@bib@innerbib\@empty
\bibitem [{\citenamefont {Steier}\ \emph {et~al.}(2018)\citenamefont {Steier},
  \citenamefont {All{\'e}zy}, \citenamefont {Anders}, \citenamefont {Baptiste},
  \citenamefont {Buice}, \citenamefont {Chow}, \citenamefont {Cutler},
  \citenamefont {Donahue}, \citenamefont {Filippetto}, \citenamefont {Harkins}
  \emph {et~al.}}]{Steier2018}%
  \BibitemOpen
  \bibfield  {author} {\bibinfo {author} {\bibfnamefont {C.}~\bibnamefont
  {Steier}}, \bibinfo {author} {\bibfnamefont {A.}~\bibnamefont {All{\'e}zy}},
  \bibinfo {author} {\bibfnamefont {A.}~\bibnamefont {Anders}}, \bibinfo
  {author} {\bibfnamefont {K.}~\bibnamefont {Baptiste}}, \bibinfo {author}
  {\bibfnamefont {E.}~\bibnamefont {Buice}}, \bibinfo {author} {\bibfnamefont
  {K.}~\bibnamefont {Chow}}, \bibinfo {author} {\bibfnamefont {G.}~\bibnamefont
  {Cutler}}, \bibinfo {author} {\bibfnamefont {R.}~\bibnamefont {Donahue}},
  \bibinfo {author} {\bibfnamefont {D.}~\bibnamefont {Filippetto}}, \bibinfo
  {author} {\bibfnamefont {J.}~\bibnamefont {Harkins}},  \emph {et~al.},\
  }\bibfield  {title} {\enquote {\bibinfo {title} {Status of the conceptual
  design of als-u},}\ }in\ \href@noop {} {\emph {\bibinfo {booktitle} {9th
  International Particle Accelerator Conference}}}\ (\bibinfo {year} {2018})\
  pp.\ \bibinfo {pages} {4134--4137}\BibitemShut {NoStop}%
\bibitem [{\citenamefont {Borland}\ \emph {et~al.}(2017)\citenamefont
  {Borland}, \citenamefont {Emery}, \citenamefont {Lindberg}, \citenamefont
  {Sajaev}, \citenamefont {Sun},\ and\ \citenamefont {Xiao}}]{Borland2017}%
  \BibitemOpen
  \bibfield  {author} {\bibinfo {author} {\bibfnamefont {M.}~\bibnamefont
  {Borland}}, \bibinfo {author} {\bibfnamefont {L.}~\bibnamefont {Emery}},
  \bibinfo {author} {\bibfnamefont {R.}~\bibnamefont {Lindberg}}, \bibinfo
  {author} {\bibfnamefont {V.}~\bibnamefont {Sajaev}}, \bibinfo {author}
  {\bibfnamefont {Y.}~\bibnamefont {Sun}}, \ and\ \bibinfo {author}
  {\bibfnamefont {A.}~\bibnamefont {Xiao}},\ }\bibfield  {title} {\enquote
  {\bibinfo {title} {Overview of lattice design and evaluation for the aps
  upgrade},}\ }\href@noop {} {\bibfield  {journal} {\bibinfo  {journal} {ICFA
  Beam Dynamics Newsletter}\ }\textbf {\bibinfo {volume} {71}} (\bibinfo {year}
  {2017})}\BibitemShut {NoStop}%
\bibitem [{\citenamefont {Bryant}(1975)}]{Bryant1975}%
  \BibitemOpen
  \bibfield  {author} {\bibinfo {author} {\bibfnamefont {P.}~\bibnamefont
  {Bryant}},\ }\href {https://cds.cern.ch/record/696405} {\emph {\bibinfo
  {title} {{A simple theory for betatron coupling}}}},\ \bibinfo {type} {Tech.
  Rep.}\ (\bibinfo  {institution} {CERN},\ \bibinfo {address} {Geneva},\
  \bibinfo {year} {1975})\BibitemShut {NoStop}%
\bibitem [{\citenamefont {Guignard}(1979)}]{Guignard1979}%
  \BibitemOpen
  \bibfield  {author} {\bibinfo {author} {\bibfnamefont {G.}~\bibnamefont
  {Guignard}},\ }\bibfield  {title} {\enquote {\bibinfo {title} {Linear
  coupling in storage rings with radiating particles},}\ }\href@noop {}
  {\bibfield  {journal} {\bibinfo  {journal} {CERN ISR—BOM/79-30, LEP note
  154}\ } (\bibinfo {year} {1979})}\BibitemShut {NoStop}%
\bibitem [{\citenamefont {Franchi}\ \emph {et~al.}(2011)\citenamefont
  {Franchi}, \citenamefont {Farvacque}, \citenamefont {Chavanne}, \citenamefont
  {Ewald}, \citenamefont {Nash}, \citenamefont {Scheidt},\ and\ \citenamefont
  {Tom\'as}}]{Franchi2011}%
  \BibitemOpen
  \bibfield  {author} {\bibinfo {author} {\bibfnamefont {A.}~\bibnamefont
  {Franchi}}, \bibinfo {author} {\bibfnamefont {L.}~\bibnamefont {Farvacque}},
  \bibinfo {author} {\bibfnamefont {J.}~\bibnamefont {Chavanne}}, \bibinfo
  {author} {\bibfnamefont {F.}~\bibnamefont {Ewald}}, \bibinfo {author}
  {\bibfnamefont {B.}~\bibnamefont {Nash}}, \bibinfo {author} {\bibfnamefont
  {K.}~\bibnamefont {Scheidt}}, \ and\ \bibinfo {author} {\bibfnamefont
  {R.}~\bibnamefont {Tom\'as}},\ }\bibfield  {title} {\enquote {\bibinfo
  {title} {Vertical emittance reduction and preservation in electron storage
  rings via resonance driving terms correction},}\ }\href {\doibase
  10.1103/PhysRevSTAB.14.034002} {\bibfield  {journal} {\bibinfo  {journal}
  {Phys. Rev. ST Accel. Beams}\ }\textbf {\bibinfo {volume} {14}},\ \bibinfo
  {pages} {034002} (\bibinfo {year} {2011})}\BibitemShut {NoStop}%
\bibitem [{\citenamefont {Courant}\ and\ \citenamefont
  {Snyder}(1958)}]{Courant1958}%
  \BibitemOpen
  \bibfield  {author} {\bibinfo {author} {\bibfnamefont {E.}~\bibnamefont
  {Courant}}\ and\ \bibinfo {author} {\bibfnamefont {H.}~\bibnamefont
  {Snyder}},\ }\bibfield  {title} {\enquote {\bibinfo {title} {Theory of the
  alternating-gradient synchrotron},}\ }\href@noop {} {\bibfield  {journal}
  {\bibinfo  {journal} {Annals of physics}\ }\textbf {\bibinfo {volume} {3}},\
  \bibinfo {pages} {1--48} (\bibinfo {year} {1958})}\BibitemShut {NoStop}%
\bibitem [{\citenamefont {Borchardt}\ \emph {et~al.}(1988)\citenamefont
  {Borchardt}, \citenamefont {Karantzoulis}, \citenamefont {Mais},\ and\
  \citenamefont {Ripken}}]{Borchardt1988}%
  \BibitemOpen
  \bibfield  {author} {\bibinfo {author} {\bibfnamefont {I.}~\bibnamefont
  {Borchardt}}, \bibinfo {author} {\bibfnamefont {E.}~\bibnamefont
  {Karantzoulis}}, \bibinfo {author} {\bibfnamefont {H.}~\bibnamefont {Mais}},
  \ and\ \bibinfo {author} {\bibfnamefont {G.}~\bibnamefont {Ripken}},\
  }\bibfield  {title} {\enquote {\bibinfo {title} {Calculation of beam
  envelopes in storage rings and transport systems in the presence of
  transverse space charge effects and coupling},}\ }\href@noop {} {\bibfield
  {journal} {\bibinfo  {journal} {Zeitschrift f{\"u}r Physik C Particles and
  Fields}\ }\textbf {\bibinfo {volume} {39}},\ \bibinfo {pages} {339--349}
  (\bibinfo {year} {1988})}\BibitemShut {NoStop}%
\bibitem [{\citenamefont {Willeke}\ and\ \citenamefont
  {Ripken}(1989)}]{Willeke1989}%
  \BibitemOpen
  \bibfield  {author} {\bibinfo {author} {\bibfnamefont {F.}~\bibnamefont
  {Willeke}}\ and\ \bibinfo {author} {\bibfnamefont {G.}~\bibnamefont
  {Ripken}},\ }\bibfield  {title} {\enquote {\bibinfo {title} {Methods of beam
  optics},}\ }in\ \href@noop {} {\emph {\bibinfo {booktitle} {AIP Conference
  Proceedings}}},\ Vol.\ \bibinfo {volume} {184}\ (\bibinfo {organization}
  {American Institute of Physics},\ \bibinfo {year} {1989})\ pp.\ \bibinfo
  {pages} {758--819}\BibitemShut {NoStop}%
\bibitem [{\citenamefont {Lebedev}\ and\ \citenamefont
  {Bogacz}(2010)}]{Lebedev2010}%
  \BibitemOpen
  \bibfield  {author} {\bibinfo {author} {\bibfnamefont {V.}~\bibnamefont
  {Lebedev}}\ and\ \bibinfo {author} {\bibfnamefont {S.}~\bibnamefont
  {Bogacz}},\ }\bibfield  {title} {\enquote {\bibinfo {title} {Betatron motion
  with coupling of horizontal and vertical degrees of freedom},}\ }\href@noop
  {} {\bibfield  {journal} {\bibinfo  {journal} {Journal of Instrumentation}\
  }\textbf {\bibinfo {volume} {5}},\ \bibinfo {pages} {P10010} (\bibinfo {year}
  {2010})}\BibitemShut {NoStop}%
\bibitem [{\citenamefont {Edwards}\ and\ \citenamefont
  {Teng}(1973)}]{Edwards1973}%
  \BibitemOpen
  \bibfield  {author} {\bibinfo {author} {\bibfnamefont {D.}~\bibnamefont
  {Edwards}}\ and\ \bibinfo {author} {\bibfnamefont {L.}~\bibnamefont {Teng}},\
  }\bibfield  {title} {\enquote {\bibinfo {title} {Parametrization of linear
  coupled motion in periodic systems},}\ }\href@noop {} {\bibfield  {journal}
  {\bibinfo  {journal} {IEEE Transactions on nuclear science}\ }\textbf
  {\bibinfo {volume} {20}},\ \bibinfo {pages} {885--888} (\bibinfo {year}
  {1973})}\BibitemShut {NoStop}%
\bibitem [{\citenamefont {Luo}(2004)}]{luo2004}%
  \BibitemOpen
  \bibfield  {author} {\bibinfo {author} {\bibfnamefont {Y.}~\bibnamefont
  {Luo}},\ }\bibfield  {title} {\enquote {\bibinfo {title} {Linear coupling
  parametrization in the action-angle frame},}\ }\href@noop {} {\bibfield
  {journal} {\bibinfo  {journal} {Physical Review Special Topics-Accelerators
  and Beams}\ }\textbf {\bibinfo {volume} {7}},\ \bibinfo {pages} {124001}
  (\bibinfo {year} {2004})}\BibitemShut {NoStop}%
\bibitem [{\citenamefont {Wolski}(2006)}]{wolski2006}%
  \BibitemOpen
  \bibfield  {author} {\bibinfo {author} {\bibfnamefont {A.}~\bibnamefont
  {Wolski}},\ }\bibfield  {title} {\enquote {\bibinfo {title} {Alternative
  approach to general coupled linear optics},}\ }\href {\doibase
  10.1103/PhysRevSTAB.9.024001} {\bibfield  {journal} {\bibinfo  {journal}
  {Phys. Rev. ST Accel. Beams}\ }\textbf {\bibinfo {volume} {9}},\ \bibinfo
  {pages} {024001} (\bibinfo {year} {2006})}\BibitemShut {NoStop}%
\bibitem [{\citenamefont {Guignard}(1976)}]{Guignard1976}%
  \BibitemOpen
  \bibfield  {author} {\bibinfo {author} {\bibfnamefont {G.}~\bibnamefont
  {Guignard}},\ }\bibfield  {title} {\enquote {\bibinfo {title} {The general
  theory of all sum and difference resonances in a three-dimensional magnetic
  field in a synchrotron},}\ }\href@noop {} {\  (\bibinfo {year}
  {1976})}\BibitemShut {NoStop}%
\bibitem [{\citenamefont {Chao}(1979)}]{Chao1979}%
  \BibitemOpen
  \bibfield  {author} {\bibinfo {author} {\bibfnamefont {A.}~\bibnamefont
  {Chao}},\ }\bibfield  {title} {\enquote {\bibinfo {title} {Evaluation of beam
  distribution parameters in an electron storage ring},}\ }\href@noop {}
  {\bibfield  {journal} {\bibinfo  {journal} {Journal of Applied Physics}\
  }\textbf {\bibinfo {volume} {50}},\ \bibinfo {pages} {595--598} (\bibinfo
  {year} {1979})}\BibitemShut {NoStop}%
\bibitem [{\citenamefont {Chao}(1981)}]{Chao1981}%
  \BibitemOpen
  \bibfield  {author} {\bibinfo {author} {\bibfnamefont {A.}~\bibnamefont
  {Chao}},\ }\bibfield  {title} {\enquote {\bibinfo {title} {Evaluation of
  radiative spin polarization in an electron storage ring},}\ }\href@noop {}
  {\bibfield  {journal} {\bibinfo  {journal} {Nuclear Instruments and Methods}\
  }\textbf {\bibinfo {volume} {180}},\ \bibinfo {pages} {29--36} (\bibinfo
  {year} {1981})}\BibitemShut {NoStop}%
\bibitem [{\citenamefont {Chao}(2009)}]{Chao2009}%
  \BibitemOpen
  \bibfield  {author} {\bibinfo {author} {\bibfnamefont {A.}~\bibnamefont
  {Chao}},\ }\bibfield  {title} {\enquote {\bibinfo {title} {{SLIM} -- a
  formalism for linear coupled systems},}\ }\href@noop {} {\bibfield  {journal}
  {\bibinfo  {journal} {Chinese Physics C}\ }\textbf {\bibinfo {volume} {33}},\
  \bibinfo {pages} {115} (\bibinfo {year} {2009})}\BibitemShut {NoStop}%
\bibitem [{\citenamefont {Ohmi}\ \emph {et~al.}(1994)\citenamefont {Ohmi},
  \citenamefont {Hirata},\ and\ \citenamefont {Oide}}]{Ohmi1994}%
  \BibitemOpen
  \bibfield  {author} {\bibinfo {author} {\bibfnamefont {K.}~\bibnamefont
  {Ohmi}}, \bibinfo {author} {\bibfnamefont {K.}~\bibnamefont {Hirata}}, \ and\
  \bibinfo {author} {\bibfnamefont {K.}~\bibnamefont {Oide}},\ }\bibfield
  {title} {\enquote {\bibinfo {title} {From the beam-envelope matrix to
  synchrotron-radiation integrals},}\ }\href@noop {} {\bibfield  {journal}
  {\bibinfo  {journal} {Physical Review E}\ }\textbf {\bibinfo {volume} {49}},\
  \bibinfo {pages} {751} (\bibinfo {year} {1994})}\BibitemShut {NoStop}%
\bibitem [{\citenamefont {Hirata}(1988)}]{hirata1988}%
  \BibitemOpen
  \bibfield  {author} {\bibinfo {author} {\bibfnamefont {K.}~\bibnamefont
  {Hirata}},\ }\bibfield  {title} {\enquote {\bibinfo {title} {An introduction
  to sad},}\ }\href@noop {} {\  (\bibinfo {year} {1988})}\BibitemShut {NoStop}%
\bibitem [{\citenamefont {Terebilo}(2001)}]{terebilo2001}%
  \BibitemOpen
  \bibfield  {author} {\bibinfo {author} {\bibfnamefont {A.}~\bibnamefont
  {Terebilo}},\ }\bibfield  {title} {\enquote {\bibinfo {title} {Accelerator
  modeling with matlab accelerator toolbox},}\ }in\ \href@noop {} {\emph
  {\bibinfo {booktitle} {PACS2001. Proceedings of the 2001 Particle Accelerator
  Conference (Cat. No. 01CH37268)}}},\ Vol.~\bibinfo {volume} {4}\ (\bibinfo
  {organization} {IEEE},\ \bibinfo {year} {2001})\ pp.\ \bibinfo {pages}
  {3203--3205}\BibitemShut {NoStop}%
\bibitem [{\citenamefont {Chao}()}]{ChaoNote}%
  \BibitemOpen
  \bibfield  {author} {\bibinfo {author} {\bibfnamefont {A.}~\bibnamefont
  {Chao}},\ }\href@noop {} {\enquote {\bibinfo {title} {Lecture notes on
  special topics in accelerator physics: {SLIM} formalism--orbital motion},}\
  }\bibinfo {howpublished}
  {\url{https://www.slac.stanford.edu/~achao/SLIM1.pdf}}\BibitemShut {NoStop}%
\bibitem [{mad()}]{madx}%
  \BibitemOpen
  \href@noop {} {\enquote {\bibinfo {title} {{MAD - Methodical Accelerator
  Design}},}\ }\bibinfo {howpublished}
  {\url{https://madx.web.cern.ch/madx/}}\BibitemShut {NoStop}%
\bibitem [{\citenamefont {Sands}(1970)}]{Sands1970}%
  \BibitemOpen
  \bibfield  {author} {\bibinfo {author} {\bibfnamefont {M.}~\bibnamefont
  {Sands}},\ }\bibfield  {title} {\enquote {\bibinfo {title} {Physics of
  electron storage rings: An introduction},}\ }\href@noop {} {\bibfield
  {journal} {\bibinfo  {journal} {SLAC-121}\ } (\bibinfo {year}
  {1970})}\BibitemShut {NoStop}%
\bibitem [{\citenamefont {Sagan}\ and\ \citenamefont
  {Rubin}(1999)}]{sagan1999}%
  \BibitemOpen
  \bibfield  {author} {\bibinfo {author} {\bibfnamefont {D.}~\bibnamefont
  {Sagan}}\ and\ \bibinfo {author} {\bibfnamefont {D.}~\bibnamefont {Rubin}},\
  }\bibfield  {title} {\enquote {\bibinfo {title} {Linear analysis of coupled
  lattices},}\ }\href@noop {} {\bibfield  {journal} {\bibinfo  {journal}
  {Physical Review Special Topics-Accelerators and Beams}\ }\textbf {\bibinfo
  {volume} {2}},\ \bibinfo {pages} {074001} (\bibinfo {year}
  {1999})}\BibitemShut {NoStop}%
\bibitem [{\citenamefont {Borland}(2000)}]{Borland2000}%
  \BibitemOpen
  \bibfield  {author} {\bibinfo {author} {\bibfnamefont {M.}~\bibnamefont
  {Borland}},\ }\bibfield  {title} {\enquote {\bibinfo {title} {{elegant: A
  Flexible SDDS-Compliant Code for Accelerator Simulation}},}\ }\href@noop {}
  {\bibfield  {journal} {\bibinfo  {journal} {Advanced Photon Source, LS-287}\
  } (\bibinfo {year} {2000})}\BibitemShut {NoStop}%
\bibitem [{\citenamefont {Grote}\ \emph {et~al.}(1989)\citenamefont {Grote},
  \citenamefont {Iselin}, \citenamefont {Keil},\ and\ \citenamefont
  {Niederer}}]{Grote1989}%
  \BibitemOpen
  \bibfield  {author} {\bibinfo {author} {\bibfnamefont {H.}~\bibnamefont
  {Grote}}, \bibinfo {author} {\bibfnamefont {C.}~\bibnamefont {Iselin}},
  \bibinfo {author} {\bibfnamefont {E.}~\bibnamefont {Keil}}, \ and\ \bibinfo
  {author} {\bibfnamefont {J.}~\bibnamefont {Niederer}},\ }\bibfield  {title}
  {\enquote {\bibinfo {title} {The mad program},}\ }in\ \href@noop {} {\emph
  {\bibinfo {booktitle} {Proceedings of the 1989 IEEE Particle Accelerator
  Conference}}}\ (\bibinfo {organization} {IEEE},\ \bibinfo {year} {1989})\
  pp.\ \bibinfo {pages} {1292--1294}\BibitemShut {NoStop}%
\bibitem [{\citenamefont {Yoshida}(1990)}]{Yoshida1990}%
  \BibitemOpen
  \bibfield  {author} {\bibinfo {author} {\bibfnamefont {H.}~\bibnamefont
  {Yoshida}},\ }\bibfield  {title} {\enquote {\bibinfo {title} {{Construction
  of higher order symplectic integrators}},}\ }\href {\doibase
  10.1016/0375-9601(90)90092-3} {\bibfield  {journal} {\bibinfo  {journal}
  {Phys. Lett.}\ }\textbf {\bibinfo {volume} {A150}},\ \bibinfo {pages}
  {262--268} (\bibinfo {year} {1990})}\BibitemShut {NoStop}%
\bibitem [{\citenamefont {Carey}\ \emph {et~al.}(1998)\citenamefont {Carey},
  \citenamefont {Brown},\ and\ \citenamefont {Rothacker}}]{Carey1998}%
  \BibitemOpen
  \bibfield  {author} {\bibinfo {author} {\bibfnamefont {D.}~\bibnamefont
  {Carey}}, \bibinfo {author} {\bibfnamefont {K.}~\bibnamefont {Brown}}, \ and\
  \bibinfo {author} {\bibfnamefont {F.}~\bibnamefont {Rothacker}},\ }\bibfield
  {title} {\enquote {\bibinfo {title} {{Third-Order TRANSPORT with MAD Input-A
  Computer Program for Designing Charged Particle Beam Transport Systems}},}\
  }\href@noop {} {\bibfield  {journal} {\bibinfo  {journal}
  {Fermilab-Pub-95/069, SLAC-R-95-462, UC-414}\ } (\bibinfo {year}
  {1998})}\BibitemShut {NoStop}%
\bibitem [{\citenamefont {Berz}(1988)}]{Berz1988}%
  \BibitemOpen
  \bibfield  {author} {\bibinfo {author} {\bibfnamefont {A.}~\bibnamefont
  {Berz}},\ }\bibfield  {title} {\enquote {\bibinfo {title} {Differential
  algebraic description of beam dynamics to very high orders},}\ }\href@noop {}
  {\bibfield  {journal} {\bibinfo  {journal} {Part. Accel.}\ }\textbf {\bibinfo
  {volume} {24}},\ \bibinfo {pages} {109--124} (\bibinfo {year}
  {1988})}\BibitemShut {NoStop}%
\bibitem [{\citenamefont {Lavine}\ \emph {et~al.}(1988)\citenamefont {Lavine},
  \citenamefont {Seeman}, \citenamefont {Atwood}, \citenamefont {Himel},
  \citenamefont {Petersen},\ and\ \citenamefont {Adolphsen}}]{lavine1988}%
  \BibitemOpen
  \bibfield  {author} {\bibinfo {author} {\bibfnamefont {T.}~\bibnamefont
  {Lavine}}, \bibinfo {author} {\bibfnamefont {J.}~\bibnamefont {Seeman}},
  \bibinfo {author} {\bibfnamefont {W.}~\bibnamefont {Atwood}}, \bibinfo
  {author} {\bibfnamefont {T.}~\bibnamefont {Himel}}, \bibinfo {author}
  {\bibfnamefont {A.}~\bibnamefont {Petersen}}, \ and\ \bibinfo {author}
  {\bibfnamefont {C.}~\bibnamefont {Adolphsen}},\ }\href@noop {} {\emph
  {\bibinfo {title} {Beam determination of quadrupole misalignments and beam
  position monitor biases in the SLC linac}}},\ \bibinfo {type} {Tech. Rep.}\
  (\bibinfo  {institution} {Stanford Linear Accelerator Center},\ \bibinfo
  {year} {1988})\BibitemShut {NoStop}%
\bibitem [{\citenamefont {Portmann}\ \emph {et~al.}(1995)\citenamefont
  {Portmann}, \citenamefont {Robin},\ and\ \citenamefont
  {Schachinger}}]{portmann1995}%
  \BibitemOpen
  \bibfield  {author} {\bibinfo {author} {\bibfnamefont {G.}~\bibnamefont
  {Portmann}}, \bibinfo {author} {\bibfnamefont {D.}~\bibnamefont {Robin}}, \
  and\ \bibinfo {author} {\bibfnamefont {L.}~\bibnamefont {Schachinger}},\
  }\bibfield  {title} {\enquote {\bibinfo {title} {Automated beam based
  alignment of the als quadrupoles},}\ }in\ \href@noop {} {\emph {\bibinfo
  {booktitle} {Proceedings Particle Accelerator Conference}}},\ Vol.~\bibinfo
  {volume} {4}\ (\bibinfo {organization} {IEEE},\ \bibinfo {year} {1995})\ pp.\
  \bibinfo {pages} {2693--2695}\BibitemShut {NoStop}%
\bibitem [{\citenamefont {Chao}\ and\ \citenamefont
  {Deng}(2021)}]{chao-deng2021}%
  \BibitemOpen
  \bibfield  {author} {\bibinfo {author} {\bibfnamefont {A.}~\bibnamefont
  {Chao}}\ and\ \bibinfo {author} {\bibfnamefont {X.}~\bibnamefont {Deng}},\
  }\href@noop {} {}\bibinfo {howpublished} {personal communication} (\bibinfo
  {year} {2021})\BibitemShut {NoStop}%
\bibitem [{\citenamefont {Ratner}\ and\ \citenamefont
  {Chao}(2010)}]{ratner2010}%
  \BibitemOpen
  \bibfield  {author} {\bibinfo {author} {\bibfnamefont {D.}~\bibnamefont
  {Ratner}}\ and\ \bibinfo {author} {\bibfnamefont {A.}~\bibnamefont {Chao}},\
  }\bibfield  {title} {\enquote {\bibinfo {title} {Steady-state microbunching
  in a storage ring for generating coherent radiation},}\ }\href@noop {}
  {\bibfield  {journal} {\bibinfo  {journal} {Physical review letters}\
  }\textbf {\bibinfo {volume} {105}},\ \bibinfo {pages} {154801} (\bibinfo
  {year} {2010})}\BibitemShut {NoStop}%
\bibitem [{\citenamefont {Du}\ \emph {et~al.}(2021)\citenamefont {Du},
  \citenamefont {Wang}, \citenamefont {Ji},\ and\ \citenamefont
  {Tian}}]{du2021}%
  \BibitemOpen
  \bibfield  {author} {\bibinfo {author} {\bibfnamefont {C.}~\bibnamefont
  {Du}}, \bibinfo {author} {\bibfnamefont {J.}~\bibnamefont {Wang}}, \bibinfo
  {author} {\bibfnamefont {D.}~\bibnamefont {Ji}}, \ and\ \bibinfo {author}
  {\bibfnamefont {S.}~\bibnamefont {Tian}},\ }\bibfield  {title} {\enquote
  {\bibinfo {title} {A conceptual design study of generating locally-round beam
  in a diffraction-limited storage ring using skew quadrupole triplets},}\
  }\href@noop {} {\bibfield  {journal} {\bibinfo  {journal} {Nuclear
  Instruments and Methods in Physics Research Section A: Accelerators,
  Spectrometers, Detectors and Associated Equipment}\ }\textbf {\bibinfo
  {volume} {992}},\ \bibinfo {pages} {165052} (\bibinfo {year}
  {2021})}\BibitemShut {NoStop}%
\bibitem [{\citenamefont {Safranek}(1997)}]{safranek1997}%
  \BibitemOpen
  \bibfield  {author} {\bibinfo {author} {\bibfnamefont {J.}~\bibnamefont
  {Safranek}},\ }\bibfield  {title} {\enquote {\bibinfo {title} {Experimental
  determination of storage ring optics using orbit response measurements},}\
  }\href@noop {} {\bibfield  {journal} {\bibinfo  {journal} {Nuclear
  Instruments and Methods in Physics Research Section A: Accelerators,
  Spectrometers, Detectors and Associated Equipment}\ }\textbf {\bibinfo
  {volume} {388}},\ \bibinfo {pages} {27--36} (\bibinfo {year}
  {1997})}\BibitemShut {NoStop}%
\bibitem [{\citenamefont {Borland}\ \emph {et~al.}(2009)\citenamefont
  {Borland}, \citenamefont {Sajaev}, \citenamefont {Emery},\ and\ \citenamefont
  {Xiao}}]{borland2009}%
  \BibitemOpen
  \bibfield  {author} {\bibinfo {author} {\bibfnamefont {M.}~\bibnamefont
  {Borland}}, \bibinfo {author} {\bibfnamefont {V.}~\bibnamefont {Sajaev}},
  \bibinfo {author} {\bibfnamefont {L.}~\bibnamefont {Emery}}, \ and\ \bibinfo
  {author} {\bibfnamefont {A.}~\bibnamefont {Xiao}},\ }\bibfield  {title}
  {\enquote {\bibinfo {title} {Direct methods of optimization of storage ring
  dynamic and momentum aperture},}\ }\href@noop {} {\bibfield  {journal}
  {\bibinfo  {journal} {Proceedings of PAC09, TH6PFP062}\ } (\bibinfo {year}
  {2009})}\BibitemShut {NoStop}%
\bibitem [{\citenamefont {Yang}\ \emph {et~al.}(2011)\citenamefont {Yang},
  \citenamefont {Li}, \citenamefont {Guo},\ and\ \citenamefont
  {Krinsky}}]{yang2011}%
  \BibitemOpen
  \bibfield  {author} {\bibinfo {author} {\bibfnamefont {L.}~\bibnamefont
  {Yang}}, \bibinfo {author} {\bibfnamefont {Y.}~\bibnamefont {Li}}, \bibinfo
  {author} {\bibfnamefont {W.}~\bibnamefont {Guo}}, \ and\ \bibinfo {author}
  {\bibfnamefont {S.}~\bibnamefont {Krinsky}},\ }\bibfield  {title} {\enquote
  {\bibinfo {title} {Multiobjective optimization of dynamic aperture},}\
  }\href@noop {} {\bibfield  {journal} {\bibinfo  {journal} {Physical Review
  Special Topics-Accelerators and Beams}\ }\textbf {\bibinfo {volume} {14}},\
  \bibinfo {pages} {054001} (\bibinfo {year} {2011})}\BibitemShut {NoStop}%
\bibitem [{\citenamefont {Huang}\ and\ \citenamefont
  {Safranek}(2014)}]{huang2014}%
  \BibitemOpen
  \bibfield  {author} {\bibinfo {author} {\bibfnamefont {X.}~\bibnamefont
  {Huang}}\ and\ \bibinfo {author} {\bibfnamefont {J.}~\bibnamefont
  {Safranek}},\ }\bibfield  {title} {\enquote {\bibinfo {title} {Nonlinear
  dynamics optimization with particle swarm and genetic algorithms for spear3
  emittance upgrade},}\ }\href@noop {} {\bibfield  {journal} {\bibinfo
  {journal} {Nuclear Instruments and Methods in Physics Research Section A:
  Accelerators, Spectrometers, Detectors and Associated Equipment}\ }\textbf
  {\bibinfo {volume} {757}},\ \bibinfo {pages} {48--53} (\bibinfo {year}
  {2014})}\BibitemShut {NoStop}%
\bibitem [{\citenamefont {Huang}\ and\ \citenamefont
  {Safranek}(2015)}]{huang2015}%
  \BibitemOpen
  \bibfield  {author} {\bibinfo {author} {\bibfnamefont {X.}~\bibnamefont
  {Huang}}\ and\ \bibinfo {author} {\bibfnamefont {J.}~\bibnamefont
  {Safranek}},\ }\bibfield  {title} {\enquote {\bibinfo {title} {Online
  optimization of storage ring nonlinear beam dynamics},}\ }\href@noop {}
  {\bibfield  {journal} {\bibinfo  {journal} {Physical Review Special
  Topics-Accelerators and Beams}\ }\textbf {\bibinfo {volume} {18}},\ \bibinfo
  {pages} {084001} (\bibinfo {year} {2015})}\BibitemShut {NoStop}%
\bibitem [{\citenamefont {Iselin}(1984)}]{Iselin1984}%
  \BibitemOpen
  \bibfield  {author} {\bibinfo {author} {\bibfnamefont {F.}~\bibnamefont
  {Iselin}},\ }\bibfield  {title} {\enquote {\bibinfo {title} {Lie
  transformations and transport equations for combined-function dipoles},}\
  }\href@noop {} {\bibfield  {journal} {\bibinfo  {journal} {Part. Accel.}\
  }\textbf {\bibinfo {volume} {17}},\ \bibinfo {pages} {143--155} (\bibinfo
  {year} {1984})}\BibitemShut {NoStop}%
\bibitem [{\citenamefont {Halbach}(1980)}]{halbach1980}%
  \BibitemOpen
  \bibfield  {author} {\bibinfo {author} {\bibfnamefont {K.}~\bibnamefont
  {Halbach}},\ }\bibfield  {title} {\enquote {\bibinfo {title} {Design of
  permanent multipole magnets with oriented rare earth cobalt material},}\
  }\href@noop {} {\bibfield  {journal} {\bibinfo  {journal} {Nuclear
  instruments and methods}\ }\textbf {\bibinfo {volume} {169}},\ \bibinfo
  {pages} {1--10} (\bibinfo {year} {1980})}\BibitemShut {NoStop}%
\bibitem [{\citenamefont {Smith}(1986)}]{Smith1986}%
  \BibitemOpen
  \bibfield  {author} {\bibinfo {author} {\bibfnamefont {L.}~\bibnamefont
  {Smith}},\ }\bibfield  {title} {\enquote {\bibinfo {title} {Effects of
  wigglers and undulators on beam dynamics},}\ }\href@noop {} {\bibfield
  {journal} {\bibinfo  {journal} {LBL-21391, ESG-18}\ } (\bibinfo {year}
  {1986})}\BibitemShut {NoStop}%
\bibitem [{\citenamefont {Zhang}\ \emph {et~al.}(2021)\citenamefont {Zhang},
  \citenamefont {Deng}, \citenamefont {Pan}, \citenamefont {Li}, \citenamefont
  {Zhou}, \citenamefont {Huang}, \citenamefont {Li}, \citenamefont {Tang},\
  and\ \citenamefont {Chao}}]{Zhang2021}%
  \BibitemOpen
  \bibfield  {author} {\bibinfo {author} {\bibfnamefont {Y.}~\bibnamefont
  {Zhang}}, \bibinfo {author} {\bibfnamefont {X.}~\bibnamefont {Deng}},
  \bibinfo {author} {\bibfnamefont {Z.}~\bibnamefont {Pan}}, \bibinfo {author}
  {\bibfnamefont {Z.}~\bibnamefont {Li}}, \bibinfo {author} {\bibfnamefont
  {K.}~\bibnamefont {Zhou}}, \bibinfo {author} {\bibfnamefont {W.}~\bibnamefont
  {Huang}}, \bibinfo {author} {\bibfnamefont {R.}~\bibnamefont {Li}}, \bibinfo
  {author} {\bibfnamefont {C.}~\bibnamefont {Tang}}, \ and\ \bibinfo {author}
  {\bibfnamefont {A.}~\bibnamefont {Chao}},\ }\bibfield  {title} {\enquote
  {\bibinfo {title} {Ultralow longitudinal emittance storage rings},}\
  }\href@noop {} {\bibfield  {journal} {\bibinfo  {journal} {Physical Review
  Accelerators and Beams}\ }\textbf {\bibinfo {volume} {24}},\ \bibinfo {pages}
  {090701} (\bibinfo {year} {2021})}\BibitemShut {NoStop}%
\end{thebibliography}%

\end{document}